\documentclass{webofc}

\usepackage[varg]{txfonts}   
\usepackage{hyperref}
\usepackage{url}
\usepackage[mathlines]{lineno}
\usepackage{float}

\hypersetup{colorlinks=true,citecolor=blue,urlcolor=blue,linkcolor=blue}
\begin{document}
\title{Study of baryon-strangeness and charge-strangeness correlations in Pb--Pb collisions at $\sqrt{s_\mathrm{NN}}$ = 5.02 TeV with ALICE}
%
%

\author{\firstname{Swati} \lastname{Saha}\inst{1,2}\fnsep\thanks{\email{swati.saha@cern.ch}} for the ALICE Collaboration
}

\institute{School of Physical Sciences, National Institute of Science Education and Research, Jatni-752050, Odisha, India
\and
           Homi Bhabha National Institute, Training School Complex, Anushaktinagar, Mumbai-400094, Maharastra, India
          }

\abstract{
  In the quest to unravel the mysteries of the strong force and the underlying properties of the quark-gluon plasma, the ALICE collaboration at CERN has carried out a comprehensive study focusing on the correlations between net-conserved quantities such as net-baryon, net-charge and net-strangeness. These correlations play a crucial role in the study of QCD phase structure as they are closely related to the ratios of thermodynamic susceptibilities in lattice QCD calculations. This work mainly focuses on the correlations between net-kaon and net-proton, and net-kaon and net-charge in Pb--Pb collisions at $\sqrt{s_\mathrm{NN}} = 5.02$ TeV using data recorded during LHC Run 2. The net-proton and net-kaon serve as proxies for net-baryon and net-strangeness, respectively, with measurements analyzed as a function of collision centrality. Theoretical predictions from the Thermal-FIST model are compared with experimental results, providing insights into the effects of resonance decays and charge conservation laws on the correlations.
}
\maketitle
%


\section{Introduction}
\label{intro}
Quantum Chromodynamics (QCD) predicts that, at sufficiently high energy densities, nuclear matter transitions into a state called quark-gluon plasma (QGP), where quarks and gluons are no longer confined within hadrons \cite{Intro1}. Studying fluctuations and correlations of conserved charges in heavy-ion collisions provides a powerful tool for probing the QCD phase transition and the properties of the QGP medium \cite{Motiv1, Motiv2}. These observables are linked to thermodynamic susceptibilities in lattice QCD (LQCD) calculations and exhibit characteristic changes in the crossover region between low- and high-temperature phases of QCD \cite{Friman, FKarsch}. Comparing experimental measurements of these fluctuations and correlations with Hadron Resonance Gas (HRG) model predictions can enhance our understanding of freeze-out parameters and the thermodynamic properties of the medium formed in heavy-ion collisions \cite{FKarsch}. However, it is important to recognize that, in addition to critical signals related to phase transition, other dynamical sources can also influence the measurements of conserved charge fluctuations and correlations, such as quantum number conservation \cite{Qua}, volume fluctuations \cite{Vol2, Vol}, resonance decays \cite{Reso2, Reso}, etc. Understanding the effects of these individual sources is crucial. This report will primarily focus on the impact of exact conservation laws--baryon number (B), electric charge (Q), and strangeness (S)--on particle production, as well as on the role of resonance decays and final-state interactions in affecting the measurements.

For a system with pressure $P$ and temperature $T$, the thermodynamic susceptibilities of conserved charges in the framework of LQCD are obtained from partial derivatives of $P/T^{4}$ with respect to the chemical potentials of the conserved charges ($\mu_x$: $x$ = Q, B, or S):

$$\chi^{l,m,n}_\mathrm{B,Q,S} = \frac{\partial^{(l+m+n)}(P/T^{4})}{\partial^{l}(\mu_\mathrm{B}/T)\partial^{m}(\mu_\mathrm{Q}/T)\partial^{n}(\mu_\mathrm{S}/T)},$$
where $l, m, n = 1, 2, 3,..., n$ are the order of derivative. These susceptibilities are studied experimentally in terms of the ratios of cumulants of net-conserved charge distributions \cite{Sci}. Owing to the experimental challenges in detecting all baryons and strange hadrons, net-proton (p) and net-kaon (K) are employed as proxies for net-baryon and net-strangeness, respectively. The second-order ($l+m+n=2$) diagonal and off-diagonal cumulants for net-charge, net-proton, and net-kaon multiplicity distributions can be expressed as: $\sigma_{\alpha}^{2}=\langle(\delta N_\alpha)^{2}\rangle$ and $\sigma_{\alpha,\beta}^{11}=\langle(\delta N_\alpha)(\delta N_\beta)\rangle$, where $\alpha, \beta$ can be Q, p, or K, $\delta N_\alpha = (N_{\alpha^+}-N_{\alpha^-})-\langle(N_{\alpha^+}-N_{\alpha^-})\rangle$, and angular brackets $\langle...\rangle$ denote average over all events.

In this report, the cumulant ratios $C_\mathrm{p,K}=\sigma^{11}_\mathrm{p,K}/\sigma^{2}_\mathrm{K}$ and $C_\mathrm{Q,K}=\sigma^{11}_\mathrm{Q,K}/\sigma^{2}_\mathrm{K}$ (proxies of baryon-strangeness correlation and charge-strangeness correlation respectively) \cite{ObsNote1, ObsNote2} in Pb--Pb collisions at $\sqrt{s_\mathrm{NN}} = 5.02$ TeV measured using the data recorded by the ALICE \cite{Alice, Alice2} detector are presented as a function of centrality.

\section{Analysis details}
\label{expt}
The analysis is conducted using 80 million Pb--Pb collision events recorded during the 2015 LHC run, selected by a minimum-bias trigger that requires signals from both the V0A ($2.8 < \eta < 5.1$) and V0C ($-3.7 < \eta < -1.7$) detectors \cite{Alice, Alice2}. The centrality of an event is determined from the signal amplitudes in these detectors. Charged particle tracks are reconstructed in the central barrel of ALICE using the Inner Tracking System (ITS) and Time Projection Chamber (TPC) \cite{Alice, Alice2}, which provide full azimuthal coverage within the pseudorapidity range $|\eta| < 0.8$. Particle identification for pions, kaons, and protons is achieved by analyzing specific energy loss ($\mathrm{d}E/\mathrm{d}x$) in the TPC and measuring flight time from the collision's primary vertex to the Time-of-Flight (TOF) \cite{Alice, Alice2} detector. The $p_\mathrm{T}$ range for particle selection is 0.2 to 2.0 GeV/c for pions and kaons, and 0.4 to 2.0 GeV/c for protons. Measurements are corrected for detector efficiencies considering binomial response of detector as detailed in Ref.~\cite{Eff}. Track reconstruction efficiencies for protons, pions, and kaons (and their anti-particles) are determined via Monte Carlo simulations with HIJING \cite{hijing} and GEANT3 \cite{Geant}, and these efficiencies are applied to correct the diagonal and off-diagonal cumulants on a track-by-track basis. Statistical uncertainties in the measurements are estimated using the bootstrap resampling method. Systematic uncertainties are evaluated by varying the criteria for event selection, track selection, and altering the conditions used for particle identification.

\section{Results}
\label{results}

\begin{figure}[h]
\centering
\includegraphics[width=0.70\linewidth]{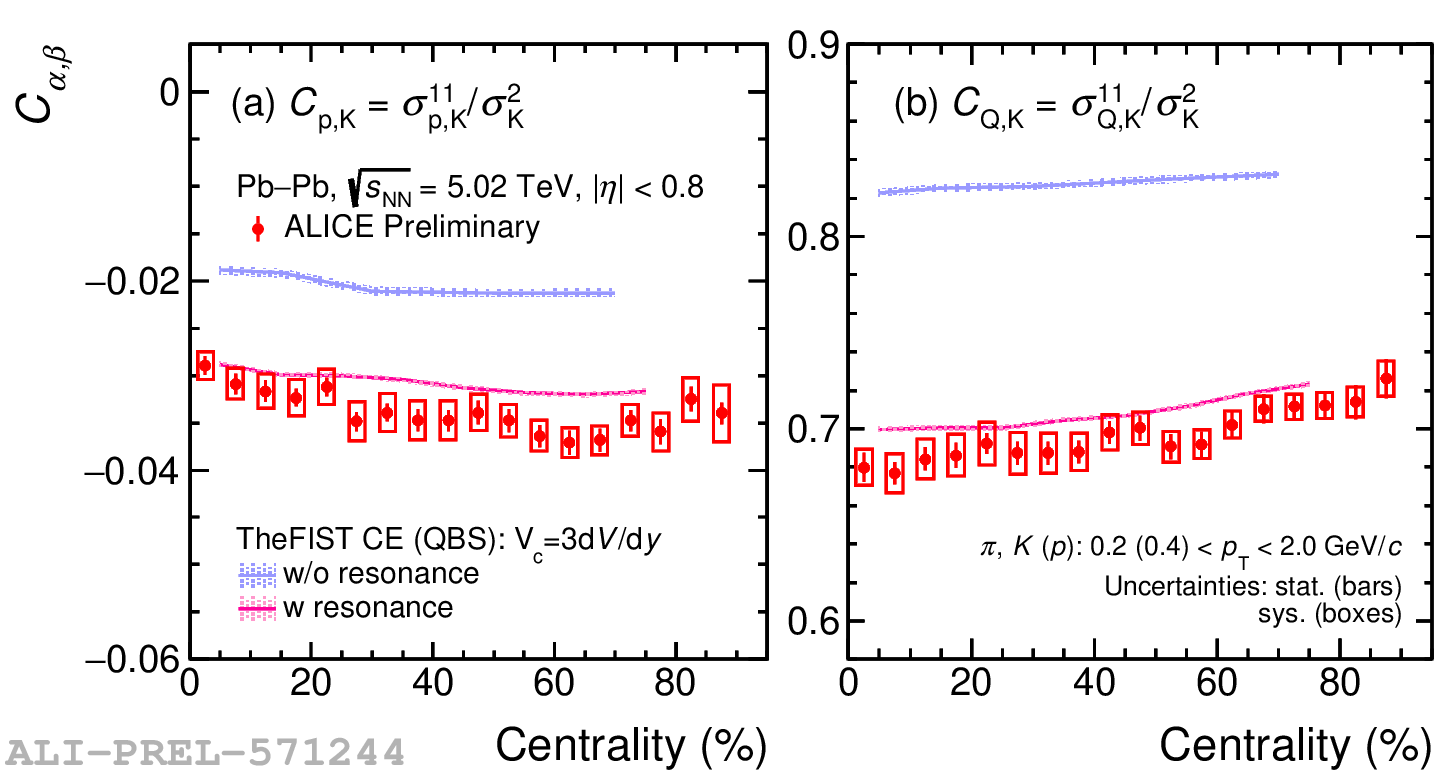}
\caption{Centrality dependence of $C_\mathrm{p,K}$ (a) and $C_\mathrm{Q,K}$ (b) in Pb--Pb collisions at $\sqrt{s_\mathrm{NN}} = 5.02$ TeV, compared to Thermal-FIST (TheFIST) results \cite{TheFIST} using the canonical ensemble (CE) formalism, enforcing exact conservation of baryon number, charge, and strangeness (B, Q, S) within a correlation volume of $V_{c} = 3\mathrm{d}V/\mathrm{d}y$. ALICE data are shown as solid red circles, while colored bands represent Thermal-FIST calculations with (w) and without (w/o) resonance contributions. The statistical (systematic) uncertainties are represented by vertical bars (boxes).}
\label{fig-1}       
\end{figure}

The measured correlation between net-proton and net-kaon, denoted as $C_\mathrm{p,K}$, is shown in the left panel of Fig.~\ref{fig-1} as a function of centrality, while the right panel displays the correlation between net-charge and net-kaon, $C_\mathrm{Q,K}$. If the event-by-event multiplicity distributions of pions, kaons, and protons follow a Poisson distribution, all the off-diagonal cumulants ($\sigma^{11}_\mathrm{p,K}$, $\sigma^{11}_\mathrm{\pi,K}$, and $\sigma^{11}_\mathrm{\pi,p}$) are zero. Consequently, the Poisson expectations for $C_\mathrm{p,K}=\sigma^{11}_\mathrm{p,K}/\sigma^{2}_\mathrm{K}$ and $C_\mathrm{Q,K} = (\sigma^{2}_\mathrm{K} + \sigma^{11}_\mathrm{p,K} + \sigma^{11}_\mathrm{\pi,K})/\sigma^{2}_\mathrm{K} = 1 + \sigma^{11}_\mathrm{p,K}/\sigma^{2}_\mathrm{K} + \sigma^{11}_\mathrm{\pi,K}/\sigma^{2}_\mathrm{K}$ would be zero and one, respectively.  Measurements of both the correlations, $C_\mathrm{p,K}$ and $C_\mathrm{Q,K}$ are found to deviate from their corresponding Poisson baselines. The effects of resonance decays on these correlations are explored by comparing the experimental measurements with Thermal-FIST model \cite{TheFIST} calculations, both with and without the inclusion of resonance decay contributions. The model calculations use the canonical ensemble (CE) formalism, which enforces exact conservation of B, Q, and S in a correlation volume of $V_c$ = $3\mathrm{d}V/\mathrm{d}y$ ($\mathrm{d}V/\mathrm{d}y$ represents volume per unit rapidity), with model parameters sourced from Ref.~\cite{TheFISTparam}. The correlations $C_\mathrm{p,K}$ and $C_\mathrm{Q,K}$ are notably enhanced across all centralities when including resonance decays, and the model results incorporating these decay contributions show better alignment with the experimental data.

\begin{figure*}[t]
\centering
\includegraphics[width=0.49\linewidth,clip]{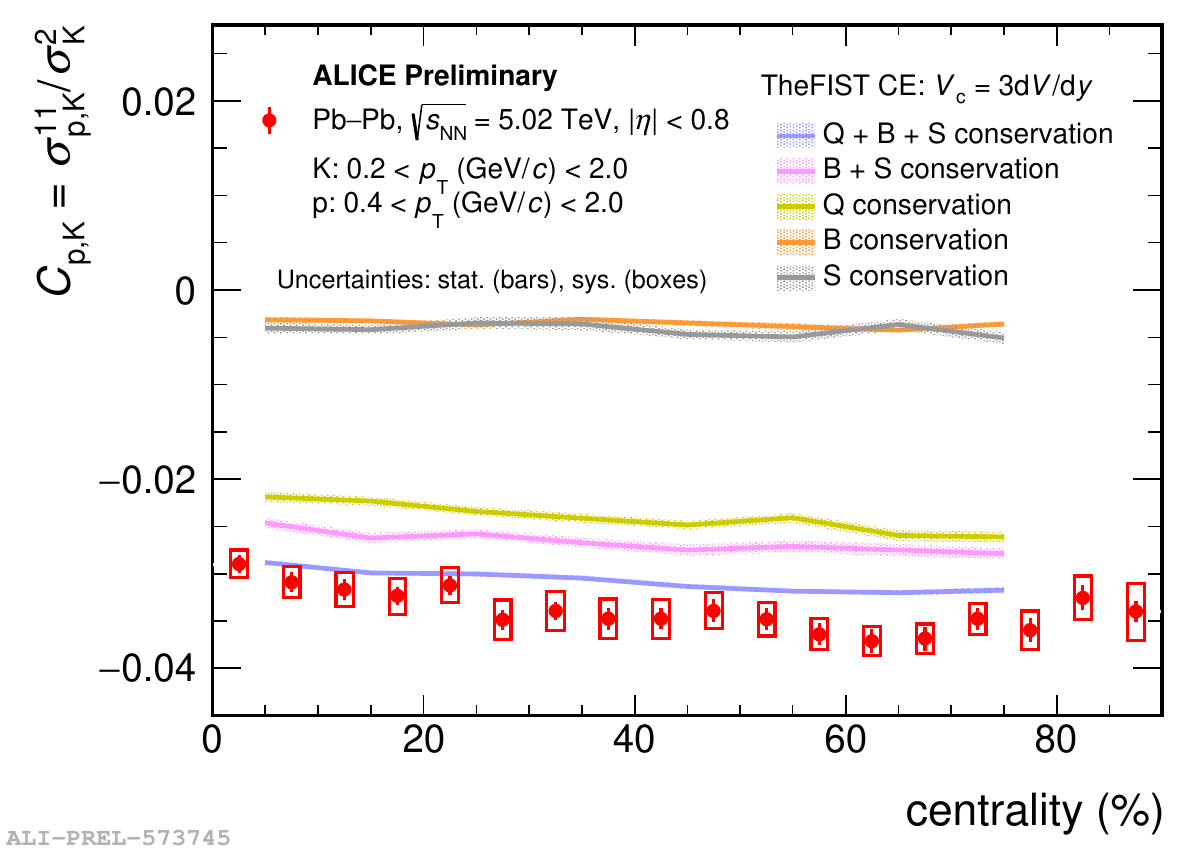}
\includegraphics[width=0.49\linewidth,clip]{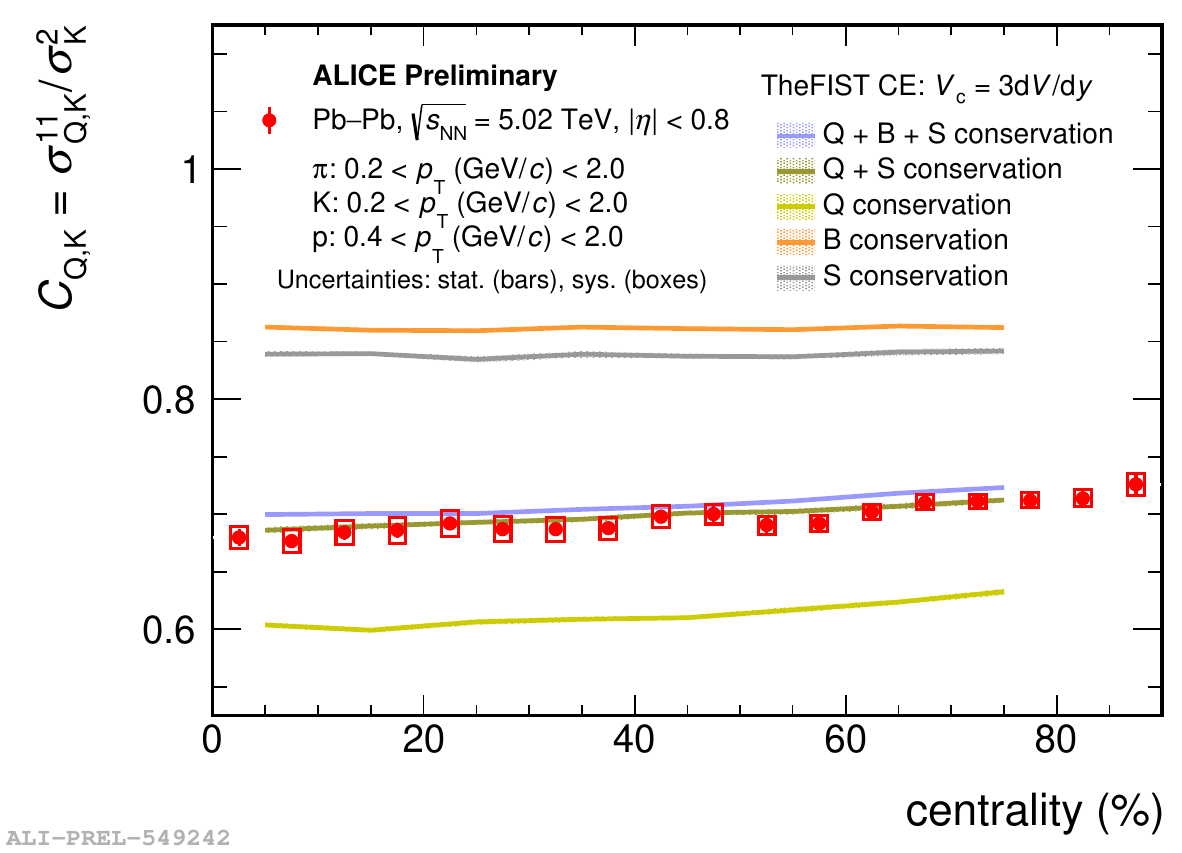}
\caption{Centrality dependence of $C_\mathrm{p,K}$ (left) and $C_\mathrm{Q,K}$ (right) in Pb--Pb collisions at $\sqrt{s_\mathrm{NN}} = 5.02$ TeV. ALICE measurements (solid red circles) are compared to predictions from the Thermal-FIST (TheFIST) \cite{TheFIST} model in the canonical ensemble (CE), accounting for different conservation effects of charge (Q), baryon number (B), and strangeness (S). The statistical (systematic) uncertainties are represented by vertical bars (boxes).}
\label{fig-2}       
\end{figure*}

The conservation laws for Q, B, and S significantly influence the observed correlations in high-energy nuclear collisions. These laws constrain the types of particles produced, leading to distinct patterns in final-state particle correlations. The effect of different charge-conservation scenarios, investigated in the framework of Thermal-FIST model is reported here. In the left (right) panel of Fig.~\ref{fig-2}, the experimental results for $C_\mathrm{p,K}$ ($C_\mathrm{Q,K}$) are compared with model predictions imposing the conservation of electric-charge-only, strangeness-only, baryon-number-only, both strangeness and baryon number (electric charge), and all three combined. $C_\mathrm{Q,K}$ is found to be mainly sensitive to conservation of both electric charge and strangeness. For $C_\mathrm{p,K}$, conservation of electric-charge-only, and both strangeness and baryon number have a similar effect. The mesaurements for both the correlations are best described by model calculations that account for the conservation of all three quantum numbers.

\section{Conclusions}
\label{conclusion}
In conclusion, the measured correlations between net-proton and net-kaon ($C_\mathrm{p,K}$) and between net-charge and net-kaon ($C_\mathrm{Q,K}$) in Pb--Pb collisions at $\sqrt{s_\mathrm{NN}} = 5.02$ TeV deviate significantly from their respective Poisson baseline values across all centralities. Resonance decays play a major role in enhancing these correlations, as demonstrated by the close agreement between experimental data and model calculations incorporating such contributions. The analysis using the Thermal-FIST model, which enforces the exact conservation of baryon number, electric charge, and strangeness, highlights the significant influence of these conservation laws on the observed correlations. Overall, the presented observables are best described when the model incorporates all three quantum number conservation laws.

\end{document}